\documentstyle[twoside,fleqn,espcrc2]{article}

\newcommand{\sbra}[1] { \left( #1 \right)}
\newcommand{\mbra}[1] { \left\{ #1 \right\}}
\newcommand{\lbra}[1] { \left[ #1 \right]}

\newcommand{\bra}[1] { \left\| #1 \right\|}

\newcommand{\AmS}{{\protect\the\textfont2
  A\kern-.1667em\lower.5ex\hbox{M}\kern-.125emS}}

\title{
Various representations of infrared
       effective lattice QCD
}
\author{
\thanks{ E-mail address:kato@hep.s.kanazawa-u.ac.jp }
\underline{S.Kato}
\address{Department of Physics, Kanazawa University Kanazawa 920-11, Japan},
M.N.Chernodub
\address{ITEP, B.Cheremushkinskaya 25, Moscow 117259, Russia},
S.Kitahara
\address{Jumonji University, Niiza, Saitama 352, Japan},
N.Nakamura \hspace{-2mm}\addtocounter{address}{-2}
\addressmark\hspace{2mm}  \\
M.I.Polikarpov \hspace{-2mm}\addtocounter{address}{-2}
\addressmark\hspace{2mm}
and \hspace{2mm}
T.Suzuki \hspace{-2mm}\addtocounter{address}{0}
\addressmark\hspace{2mm}
}

\begin{document}

\begin{abstract}
We study various representations of the infrared effective
theory of SU(2) gluodynamics starting from the monopole action
derived recently.
 We determine the coupling constants in the abelian-Higgs
model directly from lattice QCD and evaluate the type of the QCD vacuum.
The string action is derived using the BKT transformation
on the lattice. At the classical
level this action reproduces the physical string tension with a
good accuracy.
\end{abstract}

\maketitle

\input epsf

\section{INTRODUCTION}
The infrared effective theory of QCD is important for
the analytical understanding of  hadron physics. Abelian monopoles
which appear after
abelian projection of QCD \cite{'thooft} seem to be relevant
dynamical degrees of freedom for infrared region \cite{domi}.
Shiba and Suzuki \cite{shiba_suzuki} derived the monopole
action from vacuum configurations obtained in Monte-Carlo simulations
extending the method developed by Swendsen. The monopole action
was a quadratic form of monopole currents, and was not
discrete rotational invariant.

In Section~2, we present new results of a discrete rotational
invariant monopole action, the renormalized couplings are determined
by the Svendsen method, the monopoles being extracted in the
maximal abelian projection. From this monopole action
we derive the dual abelian-Higgs model (Section~3) and the string
model (Section~4) as an effective theory for SU(2) gluodynamics.

\section{MONOPOLE ACTION}

We study a discrete rotational invariant action which
includs 2-, 4- and 6-point interactions:
{\footnotesize
\begin{eqnarray*}
\!\!\!&& {\cal Z}^{mon} = \sum_{{{}^*k \in Z} \atop {\delta {}^*k=0}}
      \exp \mbra{-S^{mon}}  \\
&& S^{mon}[{}^*k] = \frac{g_m^2(b)}{2}\sbra{{}^*k,\Delta^{-1}{}^*k}
      +p(b)\bra{{}^*k}^2  \\
&+& q(b)\sum_x\sbra{\sum_{\mu=-D}^{D}{}^*k_{x,\mu}^2}^2   \\
&+& r(b)\sum_x\sum_{\mu,\nu,\delta=-D}^{D}
      {}^*k_{x,\mu}^2{}^*k_{x,\nu}^2
      \sbra{3{}^*k_{x,\delta}^2+{}^*k_{x+\nu,\delta}^2} \\
&+& \sum_i f_{i} S_2^{(i)}[{}^*k] ,
\end{eqnarray*}
}
the closed monopole currents \( ^*k\)  are defined on the dual lattice,
\(\Delta^{-1}\) is the lattice Coulomb propagator,
\(S_2^{(i)}[k]\)'s are additional quadratic forms of monopole currents
which are introduced to check whether there are any corrections
to the Coulomb interaction.
It occurs that these corrections are negligibly small for
the infrared region,
\(b(\beta,n)=na\sbra{\beta}
=\sqrt{\tilde{\kappa}(\beta,n)/\kappa_{phys}}\)
is the physical length in unit of the physical string tension
\(\kappa_{phys}\). The dimensionless string tension
\(\tilde{\kappa}\) is determined by the lattice Monte-Carlo
simulation, \(a(\beta)\) is the lattice spacing and \(n\) is the
monopole extension.

We determine the coefficients \(g_m^2(b)/2\), \(p(b)\),
\(q(b)\), \(r(b)\) numerically using the extended Swendsen method
\cite{nakam}. It turns out that these
couplings depend only on \(b\) and the
scaled action almost lies on the renormalized trajectory.
The best fit for the renormalized couplings gives:
{\footnotesize
\begin{eqnarray*}
p(b) &=& \frac{0.53(1)}{b^{3.56(4)}} ,\\
\frac{g_m^2(b)}{2}&=& 0.26(3)\sbra{\frac{4\pi}{g\sbra{b}}}^2 , \\
q(b) &=& \frac{.018(5)}{b^{6.08(4)}} ,\\
r(b) &=& \frac{.179(14) \times 10^{-3}}{b^{11.40(16)}},
\end{eqnarray*}
}

\hspace*{-4.5mm}
where
{\footnotesize
\begin{eqnarray*}
&& g^{-2}(b) =
    \frac{11}{24\pi^2}\ln\sbra{1+\frac{1}{\sbra{b\Lambda}^2}} \\
&+& \frac{17}{44\pi^2}\ln\lbra{1+\ln\sbra{1+\frac{1}
    {\sbra{b\Lambda}^2}}} ,\,\, \Lambda=0.75(5).
\end{eqnarray*}
}

\hspace*{-4.5mm}
The dependence of \(g\) on \(b\)
is consistent with the 2-loop running coupling for the small
\(b\) region and reproduces the experimental power behavior
for the large \(b\) region.

\section{DUAL ABELIAN- HIGGS MODEL FROM MONOPOLE ACTION}
The partition function of the dual abelian-Higgs model is
\begin{eqnarray*}
\!\!\! && \!\! {\cal Z}^{A.H.} \!\! = \!\! \int_{-\infty}^{+\infty}
        {\cal D}\theta \int_{-\infty}^{+\infty}{\cal D}\phi
        \exp\mbra{-{\cal S}^{A.H.}} , \\
&& {\cal S}^{A.H.} [\theta ,\phi] \!\! = \!\! \sum_P
        \frac{\beta}{2}{\theta}_p^2   \\
&+& \!\! \lambda \sum_n \sbra{{\phi}^{*}(n)\phi (n)-1}^2
        + \sum_n {\phi}^{*}(n)\phi (n)  \\
&-& \!\! \gamma \sum_{n,\mu}
        \sbra{{\phi}^{*}(n)U_{\mu}(n)\phi (n+\mu) + h.c.},
\end{eqnarray*}
where
\(U_\mu=\exp \sbra{i \theta_{\mu}}\) is the dual gauge field,
\(\theta_P \) is the field strength tensor and
\(\phi (n)=\rho_n \exp \sbra{i \varphi_n}\) is the complex Higgs field.

\vspace{3mm}
One can rewrite the above integral as the sum over the closed monopole
currents \( ^*k\) using the analogue of the BKT transformation
\cite{BKT}. The monopole action calculated in the
saddle-point approximation up to \( O(\lambda^{-2})\) terms has the form:
{\footnotesize
\begin{eqnarray*}
&&\!\!\!\!\!\!\!\!\!\!\!\!\!{\cal S}^{\rm mon}[{}^*k] =
   \frac{1}{4\beta}\sbra{{}^*k,\Delta^{-1}{}^*k} \\
+&&\!\!\!\!\!\!\!\!\!\!\!\!\!\mbra{
      \frac{1}{4 \gamma}
      +\frac{1}{\lambda}
      \sbra{\frac{15}{32 \gamma}-1}
     + \frac{1}{\lambda^2}
     \sbra{-\frac{199}{2^6 \gamma}-\frac{695}{2^7}+4\gamma}}
     \bra{{}^*k}^2  \\
+&&\!\!\!\!\!\!\!\!\!\!\!\!\!\mbra{
     -\frac{1}{2^8{\gamma}^2\lambda}
     +\frac{1}{\lambda^2}
     \sbra{\frac{71}{2^{10}\gamma}-\frac{157}{2^{12}\gamma^2}}
     }\sum_x \sbra{\sum_{\mu=-D}^{D} {}^*k_{x,\mu}^2}^2   \\
-&&\!\!\!\!\!\!\!\!\!\!\!\!\!\frac{1}{2^{11}\gamma^2\lambda^2}
     \sum_x \sum_{\mu=1}^{D} {}^*k_{x,\mu}^4
     + \frac{1}{2^{12}{\gamma}^3\lambda^2}
     \sum_x \sbra{\sum_{\mu=-D}^{D} {}^*k_{x,\mu}^2}^3\\
+&&\!\!\!\!\!\!\!\!{\cal O}\sbra{\frac{1}{\lambda^3}} .
\end{eqnarray*}
}
We consider the terms up to \(O(\lambda^{-1})\) and
we determine the \(b\) dependence of the parameters  \(\beta\)
,\(\gamma\),\(\lambda\) from the monopole action obtained numerically
by the Swendsen method.


One can estimate the type of the superconductivity of the QCD vacuum
from the Ginzburg-Landau parameter \(\kappa_{GL}\) defined by
\begin{eqnarray*}
2 \kappa_{GL}^2 = \frac{m_H^2}{m_A^2}=\frac{4\beta \lambda}{\gamma^2} .
\end{eqnarray*}
\begin{figure}
\epsfxsize=75mm\epsfysize=70mm
 \vspace{-20pt}
 \begin{center}
 \leavevmode
\epsfbox{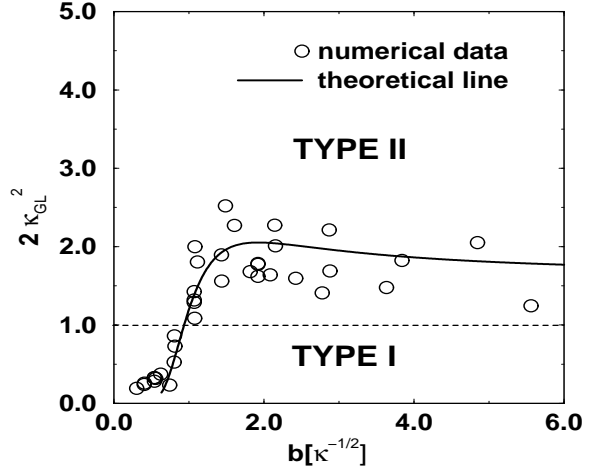}
 \end{center}
 \vspace{-37pt}
\caption{The \(b\) dependence of the Ginzburg-Landau parameter for
the QCD vacuum. The dotted
line denotes critical line between type-I and type-II superconductivity.
\label{GL}
}
\vspace{-15pt}
\end{figure}
It occurs that the QCD vacuum is a type-II superconductor
for \(b \geq 1.0 \lbra{\kappa^{-1/2}}\) (see Figure \ref{GL}).

\section{LATTICE QCD STRING}

Using the transformation suggested in Ref.~\cite{PWZ}
one can also get the effective QCD string model from the monopole
partition function:
\begin{eqnarray*}
& & {\cal S}^{str}[\sigma] \!\! = \!\! \frac{{\pi}^2}{p}
     \sbra{\sigma ,\frac{1}{\Delta + M^2}\sigma} \\
&-& \!\! \frac{{\pi}^2}{8p\tilde{\lambda} M^2}
     \Bigl(\sigma ,\frac{\Delta\sbra{0}+\Delta\sbra{1}}{
     {(\Delta + M^2)}^2}\sigma\Bigr) +
     {\cal O}(\frac{1}{M^4}) ,
\end{eqnarray*}
where \(M^2 = 1/4p\beta \)
and \(\tilde{\lambda} = p^2/16q \).The integer-valued 2-form
\(\sigma\) which is defined on the original lattice represents the
closed world surface formed by a color electric flux tube.

The leading part of this model comes from the self interaction and
from the
Coulomb interaction of the monopole action; the next-leading terms
come from the \(^*k^4\)- and \(^*k^6\)-interactions.

One can evaluate the string tension using this string model.
The Wilson loop can be written as follows:
\begin{eqnarray*}
{\langle W  \rangle}_m \!\! = \!\! \frac{1}{Z}
            \sum_{{\sigma(c_2)\in Z
            \atop{\delta \sigma=J}}}
            \exp \mbra{-S^{str} \lbra{\sigma}} ,
\end{eqnarray*}
where \(J\) is the rectangular source of a quark-antiquark pair.

In the leading order the classical string tension is:
\begin{eqnarray*}
\kappa_{th} = \frac{{\pi}^2}{b^2 p\sbra{\Delta +M^2}}\sbra{0}.
\end{eqnarray*}
 We calculate it numerically up to \(M^{-2}\)
order (\(M^2 \gg 1\) in the large \(b\) region).
\begin{figure}
\epsfxsize=75mm\epsfysize=70mm
 \vspace{-20pt}
 \begin{center}
 \leavevmode
\epsfbox{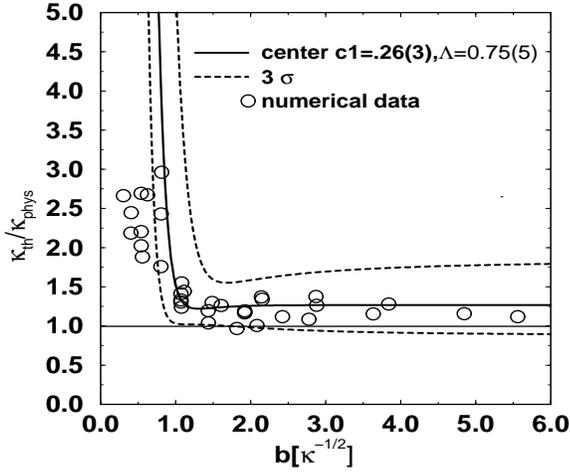}
 \end{center}
 \vspace{-37pt}
\caption{The ratio of theoretical and experimental string tension.
\label{sigma}
}
\vspace{-15pt}
\end{figure}
For \(b \geq 1.0 \lbra{\kappa^{-1/2}}\) the theoretical
string tension is almost constant and reproduces the physical
string tension fairly well (see Figure \ref{sigma}).

\section{CONCLUSIONS}

\begin{itemize}
 \item  We got the dual Ginzburg-Landau type (abelian-Higgs)
        theory, its parameters are fixed directly from the lattice QCD.
        The QCD vacuum is similar to the type-II dual superconductor.

 \item  From the effective abelian-Higgs theory we derive
        the effective string model in which the classical string
        tension almost
        coincides with the physical string tension in the infrared region.
        The difference between theoretical and experimental string
	tensions may be due to quantum corrections.

 \item The obtained QCD string model can be rewritten as follows:
\begin{eqnarray*}
{\cal S}^{str} \!\! = \!\! 4\pi^2 \beta \bra{\sigma}^2
                   -\frac{4\pi^2 \beta}{M^2}
                     \sbra{\sigma ,\Delta \sigma}
                   +{\cal O}(\frac{1}{M^4})
\end{eqnarray*}
The leading part of this action corresponds to the Nambu-Goto
action in the continuum theory.
The non-leading terms probably solve the serious problems of the simple
string model such as
the critical dimension, because the original QCD has no
such problems.

\end{itemize}

\end{document}